\documentclass[prd,showpacs,preprint,preprintnumbers,nobibnotes,superscriptaddress,floatfix]{revtex4}
\usepackage{amsmath,amssymb,amsbsy,color}
\usepackage{epsfig}
\usepackage{psfrag}

\begin{document}
\preprint{OU-HET-742}

\title{Chiral symmetry breaking in lattice QED model with fermion brane}

\author{E. Shintani}
\affiliation{
  RIKEN-BNL Research Center, Brookhaven National Laboratory, Upton, NY 11973, USA
}

\author{T. Onogi}
\affiliation{
  Department of Physics, Osaka University, Toyonaka 560-0043, Japan
}

\begin{abstract}
We propose a novel approach to the Graphene system 
using a local field theory of 4 dimensional QED model coupled to 2+1 dimensional 
Dirac fermions, whose velocity is much smaller than the speed of light.
Performing hybrid Monte Carlo simulations of this model on the lattice, 
we compute the chiral condensate and its susceptibility with 
different coupling constant, velocity parameter and flavor number.
We find that the chiral symmetry is dynamically broken in the small velocity regime
and obtain a qualitatively consistent behavior with the prediction from Schwinger-Dyson 
equations.
\end{abstract}
\maketitle

\section{Introduction}
The graphene material which is composed of two-dimensional carbon atomic seat
has attracted much attention from its specific property \cite{Novoselov:2004}.
The high conductivity of Graphene has understood by the analogical picture to
``relativistic'' massless Dirac fermion (quasiparticle) bounded in 2+1 dimension
which has been derived in based on 
tight-binding (TB) model at low energy \cite{Wallace:1947,Semenoff:1984}. 
In the effective model with massless Dirac fermion, compared with
quantum electromagnetic dynamics (QED) in particle physics, the velocity of 
fermion is 2-orders magnitude smaller than the speed of light \cite{Wallace:1947,Semenoff:1984}. 
Recent experiments suggest that this magnitude may be changed by a factor \cite{Elias:2011} 
due to significant quantum corrections. 
The spontaneous gap generation at semimetal-insulator transition point
(Mott transition) of Graphene has been predicted 
by Schwinger-Dyson equation and large N model 
\cite{Khveshchenko:2001-2004,Gorbar:2001,Gorbar:2002,Gamayun:2010,Vafek:2007,Son:2007} 
in the presence of Coulomb interaction between Dirac fermion 
before Graphene confirmed in experiments so far, however 
experimentally there has been no conclusive evidence that semimetal-insulator transition
of suspended monolayer Graphene occurs in low temperature even in 20 K \cite{Mayorov:2011}
(as opposed to this bilayer Graphene posses energy gap structure
\cite{Velasco:2012,Freitag:2012}).
It may be that the model is missing an important feature of the 
dynamics  or that the approximations used for  the theoretical prediction 
lead a wrong conclusion. In order to clarify this point 
it is necessary to carry out an exact calculation based on a solid 
theoretical approach.  

The above effective Dirac fermion system possesses chiral symmetry which could be spontaneously broken 
by strong effective Coulomb interaction due to enlarged effective fine structure constant 
$\alpha_e = e^2/(4\pi v_F)\simeq 2.2$ at small velocity $v_F\simeq 1/300$
estimated as $v_F=3 ta /2$ \cite{Wallace:1947} in the natural unit. 
Here, $t$ denotes leading hopping parameter 
of TB model and $a$ denotes distance among the nearest-neighboring carbon atoms.
The semimetal-insulator transition of Graphene 
is considered as a consequence of the energy gap generation
in the phase of chiral symmetry breaking ($\chi$SB). 
This is analogous with the  mass-gap generation in 2+1 dimension QED 
system suggested in 1/N expansion \cite{Pisarski:1984dj}, Schwinger-Dyson equation
\cite{Appelquist:1986} 20 years ago and also 
several attempts with Monte-Carlo simulation in non-Compact lattice QED
\cite{Dagotto:1988id,Hands:2002dv,Hands:2004}.
However, the present system possesses crucially different points 
from the 2+1 dimension QED system. Firstly the quasiparticle has different velocity
from speed of light in spatial 2 dimension brane and secondly there is Coulomb interaction
in spatial 3 dimension but not in spatial 2 dimension. 
The condensate for scalar bilinear operator $\langle\bar\psi\psi\rangle$ is also
as an order parameter of $\chi$SB at low temperature in this QED system. 

Note that ``chiral'' symmetry in 2+1 dimension system is defined 
for the four component Dirac spinor.
According to reference \cite{Gusynin:2007}, four component Dirac spinor is 
conventionally described as
$\psi=(\psi_\sigma^{A+},\psi_\sigma^{B+},\psi_\sigma^{A-},\psi_\sigma^{B-})$
where $\psi^{A,B\pm}$ denotes electron wave function (Bloch state)
located on sublattices (A,B) in hexagonal Graphene lattice
and two independent energy grand points (valleys) ($\pm$).
The subscript means the spin indices of electron. 
In this notation the chirality of quasiparticle corresponds to
valley indices which can be distinguished by $\gamma_5$ projection 
as well as the particle physics.
The gamma matrix is given by tensor structure as
$\gamma_0 = \sigma_0\otimes I_{2\times 2}$, $\gamma_i = -i\sigma_2\otimes \sigma_i$
($i=1,2,3$) and $\gamma_5 = i\gamma_0\gamma_1\gamma_2\gamma_3$
constructed by spin and sublattice-valley indices.
In the case of monolayer Graphene,
chiral symmetry can be regarded as global ``flavor`` $U(4)$ symmetry of quasiparticle
whose generator is given by 
$\{1,\gamma_5,i\gamma_3,[\gamma_3,\gamma_5]/2\}\otimes\sigma_{i=0,1,2,3}/2$ \cite{Gusynin:2007}
in which the first group generates the rotation operator for sublattice 
and valley points and second one generates spin rotation
(in general multilayer Graphene is described 
in $U(2N_f)$ symmetry by taking account of layer number $N_{\rm layer}$ 
as flavor $N_f=N_{\rm spin}\times N_{\rm layer}$ assuming there is no interaction 
between layers).
Using the electron wave function the chiral condensate can be expanded into 
\begin{equation}
  \langle \bar\psi\psi\rangle 
  = \sum_\sigma \int \Big[ \psi^{A+\,\dag}\psi^{B-} + \bar\psi^{A-\,\dag}\psi^{B+} +h.c. \Big]
\end{equation}
which is indeed mixture of different chirality of Graphene \cite{Gusynin:2007}.

In the leading order of extended TB model into ``non-relativistic'' QED including 
Coulomb interaction between quasiparticle, there have been many model calculations
of chiral condensate near critical point.
For example, the Schwinger-Dyson equation 
predicts that there is a critical coupling $\alpha_c$ 
above which the system has a nonzero condensate. 
In the mean-field approximation, 
$\alpha_c$ is predicted to be $\alpha_c=0.5$ \cite{Wang:2010}. 
When one-loop vacuum polarization 
effect is taken into account, the gap equation provides  stronger critical 
coupling $\alpha_c=0.92$ \cite{Gamayun:2010}.
The similar critical value has been also obtained by lattice calculation 
in the Monte-Carlo simulation \cite{Drut:2009,Armour:2010} 
or strong coupling expansion \cite{Araki:2010}
using ``non-relativistic'' QED action.

In this paper we propose a new strategy 
for non-perturbative study of the semimetal-insulator transition of 
Graphene with QED model.
The important difference from the framework of ``non-relativistic'' QED is that
we consider QED action manifestly including both 
gauge invariant interaction and velocity contribution 
in the framework of local field theory.
In our realistic setup it is possible to rigorously investigate the velocity 
contribution to $\chi$SB phenomena and compare to actual Graphene. 
We numerically show that critical behavior when changing velocity and flavor number 
is qualitatively consistent with Schwinger-Dyson equation. 

This paper is organized as the follows: 
In section \ref{sec:1} we explain the model of relativistic QED action 
with fermion brane and in section \ref{sec:sim_param}
we show the detail of setup of our lattice action and simulation. 
In section \ref{sec:res_lat} we show the behavior of chiral condensate and 
susceptibility when changing the parameters of coupling constant, fermion mass, 
velocity and flavor number.
In section \ref{sec:discuss} we summarize and discuss the future works.

\section{Modeling the relativistic QED}\label{sec:1}
In the graphene system, since the velocity of the effective Dirac fermion is extremely small, 
the following action with only the instantaneous Coulomb interaction has been
adopted \cite{Gonzalez:1999}: 
\begin{eqnarray}
  S_{\rm Coulomb} 
&=& \int dtd^2x\,\bar\psi \Big[ i\partial_t \gamma_t 
   + i v_F( \partial_x\gamma_x
   + i\partial_y\gamma_y)\Big] \psi\nonumber\\
&+& \int dt d^2x dt^\prime d^2x^\prime \bar\psi \gamma_t \psi (t,x) 
  \frac{e^2\delta(t-t^\prime)}{8\pi |x-x^\prime|} \bar\psi \gamma_t \psi (t,x^\prime).
  \label{eq:S_Coulomb}
\end{eqnarray}
Introducing the scalar potential $\phi$ as an auxiliary field, one could 
equivalently describe the action as
\begin{equation}
  S_{\rm NR} = \frac{1}{2e^2}\int dtd^3x \vec E^2
     + \int dtd^2x\,\bar\psi \Big[ i(\partial_t + i\phi) \gamma_t + i v_F(\partial_x\gamma_x
     + \partial_y\gamma_y)\Big] \psi,
  \label{eq:S_NR0}
\end{equation}
where $E \equiv - \nabla\phi$.

QED action with fermion bounded on 2+1 dimensional ``brane''
under gauge invariance in the continuum theory 
is straightforwardly written as
\begin{equation}
  S = \frac{\beta}{2}\int dtd^3x (\vec E^2 + \vec B^2)
    + \int dtd^2x\,\bar\psi\big[ iD_t \gamma_t + iv(D_x\gamma_y+D_y\gamma_y)\big]\psi
  \label{eq:S_QED}
\end{equation}
with fermi-velocity $v$ as a coefficient of gauge covariant derivative 
$D_i=\partial_i+iA_i$ and coupling constant $\beta=1/e^2$
(here we distinguish velocity parameter ``$v$'' from 
renormalized (physical) velocity ``$v_F$'' as explained below.). 
Electric and magnetic field can be described as $\vec E_i=F_{it}$, 
$\vec B_i=\varepsilon_{ijk}F_{jk}/2$ ($i=x,y,z$) with field strength 
$F_{\mu\nu}=\partial_\mu A_\nu-\partial_\nu A_\mu$, which is defined on
3+1 dimension, while fermion field $\psi$ is bounded on 2+1 dimension.
Second using simultaneous scale transformation for temporal 
coordinate and temporal gauge field normalization, 
\begin{equation}
  t \rightarrow t/v,\quad A_0 \rightarrow A_0 v,
  \label{eq:rescale} 
\end{equation}
Eq.(\ref{eq:S_QED}) is represented as
\begin{eqnarray}
  S &=& \frac{\beta}{2}\int dtd^3x( v \vec E^2 + v^{-1}\vec B^2)
     + \int dtd^2x\,\bar\psi \sum_{\mu=t,x,y}iD_\mu\gamma_{\mu} \psi,
  \label{eq:S_resQED}
\end{eqnarray}
thus fermi-velocity shows up in the anisotropy of gauge coupling. 
As a consequence of rescaling we can equivalently 
deal with kinematic term of massless fermion field as the standard QED.
Adopting the rescaled photon field for spatial direction $A_i\rightarrow A_iv$,
Eq.(\ref{eq:S_resQED}) can be described as
\begin{eqnarray}
  S &=& \frac{v\beta}{2}\int dtd^3x
    \Big[ \sum_i(\partial_i A_0-v\partial_0 A_i)^2 
   \sum_{i,j}+ (\partial_i A_j-\partial_j A_i)^2\Big]
   \nonumber\\
   &+& \int dtd^2x\,\bar\psi \Big[ i(\partial_0 + iA_0)\gamma_t + 
       \sum_{i=x,y}i(\partial_i + ivA_i)\gamma_i \Big] \psi,
  \label{eq:S_resQED2}
\end{eqnarray}
Taking the non-relativistic limit, which is interpreted as 
limit of $v\rightarrow 0$ while $v\beta$ is fixed, the spatial
photon field can be decoupled in the path integral. 
Dropping the spatial photon term, 
non-relativistic action can be derived as
\begin{equation}
  S_{\rm NR} = \frac{v\beta}{2}\int dtd^3x \vec E^2
     + \int dtd^2x\,\bar\psi \Big[ iD_t \gamma_t + i\partial_x\gamma_x
     + i\partial_y\gamma_y\Big] \psi,
  \label{eq:S_NR}
\end{equation}
so that in this limit time-rescaled version 
($t\rightarrow t/v$ and $\phi\rightarrow v\phi$) 
of the action in Eq.(\ref{eq:S_NR0}) is reproduced. 

\begin{figure}[tb]
\begin{center}
  \includegraphics[width=80mm]{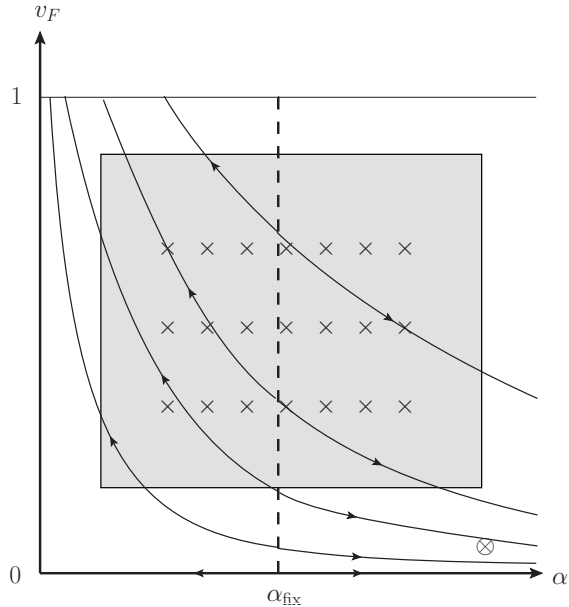}
  \caption{The sketch of renormalization flow of renormalized fermi-velocity $v_F$ and 
  effective coupling $\alpha$ with $\alpha v_F =$const.
  In the perturbation theory there is unstable infrared fixed point at $\alpha_{\rm fix}$
  which is shown as straight dashed-line. 
  Experimental point represented as cross symbol at $v_F\sim \mathcal O(10^{-2})$.
  Our lattice calculation aims to prove the wide region of phase structure 
  ($\alpha,v_F$) shown as the gray zone beyond the region perturbatively reliable.}
  \label{fig:flow}
\end{center}
\end{figure}

Here we notice that the relativistic action (\ref{eq:S_resQED}) 
is a renormalizable local quantum theory. 
Therefore, the scaling of velocity can be rigorously treated without 
the contamination from 
irrelevant operator violating gauge symmetry. 
The {\it bare}  fermi velocity $v$ has a renormalization effect due to 
quantum correction to ultraviolet divergence. 
Indeed the renormalization group study from relativistic 
perturbative calculation \cite{Gonzalez:1994,Kotov:2010}
(and also there is similar discussion in 
large N \cite{Gonzalez:2010} and renormalization group 
\cite{Juricic:2009,Juan:2010,Sinner:2010} 
in spite of non-relativistic action)
pointed out that the velocity parameter behaves a logarithmic-divergent scaling 
due to virtual fermionic loop correction. 
It turns out that the the {\it renormalized} fermi velocity $v_F$ 
has an infrared fixed point at $v_F=1$ due to the restoration of Lorentz symmetry.
Perturbative analysis has also pointed out that there is 
an infrared unstable (ultraviolet) fixed point 
of the running of fermi-velocity above the experimental one (see Figure \ref{fig:flow}).
However, as noticed in \cite{Gonzalez:1994,Kotov:2010}, 
such a fixed point is in the strong coupling region where the perturbation 
theory is not reliable.
In addition that recent experimental study using effective spectrum of 
electron-hole pair in suspended Graphene on SiO$_{2}$ wafer \cite{Elias:2011} 
also shows the effective velocity of Graphene 
around a few particle region has qualitatively similar behavior of
renormalized velocity in the perturbation theory.
Therefore, investigation of running behavior of fermi-velocity 
using non-perturbative calculation with relativistic action 
would be very important task towards more rigorous discussions than 
renormalization group study.

\section{Lattice simulation of the relativistic QED}\label{sec:sim_param}

In our Monte-Carlo study,  we employ the conventional non-compact 
gauge action 
\begin{equation}
  S_g = \sum^{\rm 4D}_x \Big[ \beta v(\partial_0 \theta_i(x) - \partial_i \theta_0(x))^2 
      + \frac{\beta}{v}(\partial_i \theta_j(x) - \partial_j \theta_i(x))^2 \Big],
\end{equation}
with anisotropic gauge coupling and vector field $\theta_\mu$.
This summation is performed in 4 dimensional coordinate space-time.
In this action there are two input parameters, $\beta$ and $v$, which correspond
to the bare gauge coupling constant and velocity. 
For the fermion action on the lattice, we use the staggered type 
\begin{eqnarray}
  S_f = \sum_{x}^{\rm 3D}\Big[ \sum_\mu\eta_\mu(x)\Big\{\bar\chi(x)U_\mu(x)\chi(x+\hat\mu)
      - U^\dag(x-\hat\mu) \chi(x-\hat\mu)\Big\} + m\bar\chi(x)\chi(x)\Big]
  \label{eq:S_f}
\end{eqnarray}
where link variable is defined as $U_\mu(x)=\exp(i\theta_\mu(x))$, and 
$\eta_\mu(x)=\Pi_{i=x_1}^{x_{\mu-1}}(-1)^i$ is Kawamoto-Smit phase factor.
In the above equation the staggered fermion is defined in 3 dimensional field, and 
thus the summation is defined in 3 dimensional coordinate space-time.
For comparison the size of z-axis, which is perpendicular coordinate for fermion brane, 
is set to same $L_z=8$ as \cite{Drut:2009}. 
Boundary condition of fermion in temporal direction is anti-periodic boundary
associated with finite temperature, and the other directions are set to periodic boundary.
In this setup gauge field propagates along z-axis without interaction of fermion.
We insert the small mass term with parameter $m$ as a probe to 
investigate chiral dynamics near chiral limit.

Conveniently staggered fermion in 3 dimensional coordinate
automatically satisfies a same flavor $U(4)$ symmetry as monolayer Graphene without 
additional counter term in the continuum limit \cite{Drut:2009}
(due to the lattice artifact it is broken to $U(1)\otimes U(1)$).
Lattice size is $30^2\times 20$, and 
we perform the Hybrid Monte Carlo (HMC) method with 
Omelyan integrator using parameter
$\lambda_c=0.1931833$ \cite{Omelyan:2002} and 
Hasenbusch accelerator \cite{Hasenbusch:2001} with mass parameter 
$m=0.05$ to generate gauge configurations with dynamical fermion.
We tune the Leapfrog time-stepping parameter $\delta_t=1/N_t$, 
where $N_t$ is number of step in unit HMC trajectory, to be that acceptance rate
is satisfied with more than 60\% in our simulation. 
From the practical point of view, 
the rescaling (\ref{eq:rescale})
has also an advantage of 
simultaneously covering the region of strong electric coupling 
and low temperature. Since temporal size is $1/v$ times larger than 
original action (\ref{eq:S_QED}), 
the lattice simulation curries out naively equivalent 
to be at $1/v$ times smaller temperature \cite{Note:1} with same computational cost. 
Note that the physical quantities obtained in this action depends on
the effective strong coupling $\alpha=1/(4\pi\beta v)$ rather than gauge coupling 
constant $\beta$ since the quantum correction from vertex with $\alpha$ is 
dominant contribution in this system if velocity is enough small. 
Correspondingly the loop correction to gauge coupling constant is higher order effect
while velocity has logarithmic divergence ($v(\mu)=v(1+\alpha/4\ln(\Lambda/\mu)$
with cut-off $\Lambda$) from at least one-loop correction
according to the perturbative analysis \cite{Gonzalez:1994}.
It turns out that observable are mostly controlled by the theoretical parameter $v$
instead of gauge coupling. 
Furthermore we introduce the Dirac ``mass'' term $m\bar\psi\psi$ which 
preserves discrete symmetry (Parity, time-reversal and charge conjugate) 
but not chiral symmetry.
Scalar condensate of quasiparticle $\langle\bar\psi\psi\rangle$
corresponds to order parameter of chiral transition, 
and associated from Goldstone theorem the massless Goldstone mode (pion) appearing into 
QED model plays a significant role of the gap generation in Graphene system
\cite{Shintani}.

\section{Result of lattice simulation}\label{sec:res_lat}
Here we investigate the chiral breaking phenomena with 
chiral condensate $\sigma = \langle \bar\chi\chi\rangle$ and 
its susceptibility $\chi_m = \partial\sigma/\partial m$ behaved 
as a function of $\beta v$, $v$ and $m$. 
The disconnected diagram appearing in chiral condensate $\sigma$ 
and chiral susceptibility $\chi_m$ is calculated by the 
noise method with 200 noise vector. 
We sample configurations at every 20 HMC trajectories, 
and employ the Jackknife error estimate with 10 bin-size.
Total statistics used in this analysis is around 600--1500 configurations,
especially more than 1000 configurations in the strong coupling region ($\beta v < 0.06$).

\begin{figure}[tb]
\begin{center}
  \includegraphics[width=80mm]{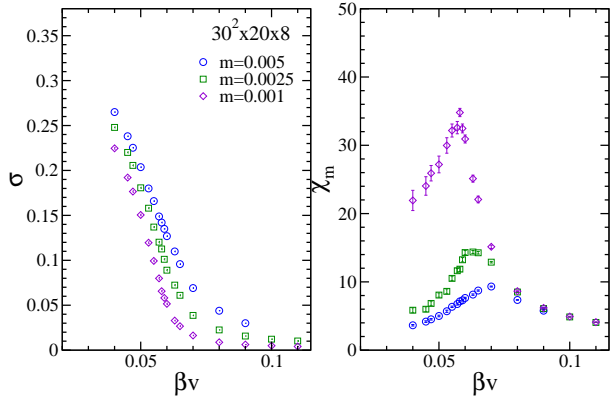}
  \caption{Chiral condensate (left) and chiral susceptibility (right) 
  of quasiparticle as a function $\beta v$ which corresponds to 
  effective coupling constant $\alpha$ at fixed velocity in $v=0.1$.}
  \label{fig:sigma_beta}
\end{center}
\end{figure}

\begin{figure}[tb]
\begin{center}
  \includegraphics[width=80mm]{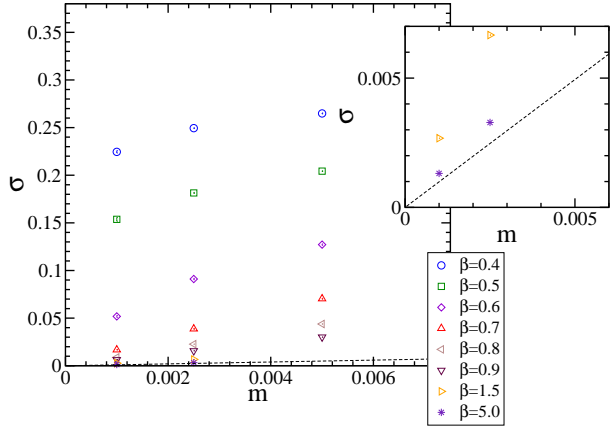}
  \caption{Mass dependence of chiral condensate at several 
  gauge coupling $\beta$ at fixed velocity in $v=0.1$.
  Linear-dashed line shows the leading perturbation result.}
  \label{fig:sigma_m}
\end{center}
\end{figure}

Chiral condensate is an order parameter of spontaneous chiral symmetry breaking in which 
the discontinuous point along the coupling constant regards as second-order critical point
in the limit of massless on infinite volume.
At the finite mass and lattice volume, if there is critical point associated 
with spontaneous chiral symmetry breaking in this system, 
chiral condensate above critical coupling $\alpha_c$ remains finite value, otherwise 
below $\alpha_c$ chiral condensate advances toward zero when approaching $m=0$. 
Figure \ref{fig:sigma_beta} clearly shows the expected critical behavior that, 
when mass parameter is decrease, $\sigma$ at weak coupling region, which is at 
the large $\beta v$, goes to zero while at strong coupling region 
$\sigma$ remains finite value, and critical coupling is found to be 
$\beta v=0.05$--$0.06$ at fixed $v=0.1$.
Chiral susceptibility also shows significant behavior around such critical point. 
When $m$ goes down close to zero, peak position of $\chi_m$ grows up and shifts 
from weak coupling to strong coupling region. 
We expect the singularity for critical phenomena at massless point 
will appear at $\beta v=$0.05--0.055, which corresponds to critical coupling 
$\alpha_c=$1.45--1.59.
This value is similar order of the prediction in gap equation 
quoted to $\alpha_c=0.92$ \cite{Gamayun:2010}
or in non-relativistic lattice calculation as $\alpha_c=1.02$ \cite{Drut:2009}.
Figure \ref{fig:sigma_m} also shows a significant
change as that below $\alpha_c$
we can see that $\sigma$ approaches to zero 
at chiral limit, especially near $\alpha_c$ its non-linear behavior,  
while above $\alpha_c$ $\sigma$ does not vanish. 
At very weak region in which the coupling anisotropy is negligible since 
$\beta$ is too large, $\sigma$ can be described as a perturbative line, 
and actually in Figure \ref{fig:sigma_m} mass dependence of 
$\sigma$ approaches to the leading 
perturbative line obtained in lattice perturbation.

\begin{figure}[tb]
\begin{center}
  \includegraphics[width=75mm]{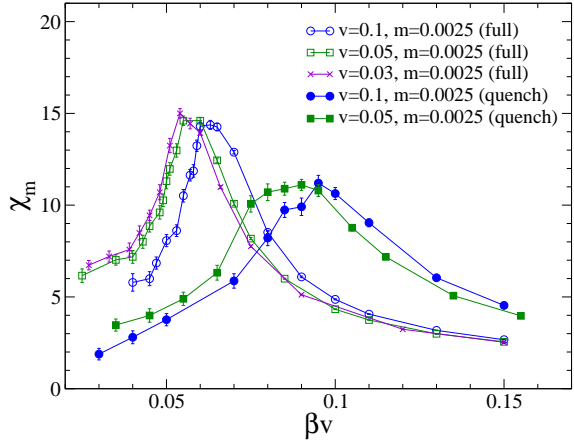}
  \caption{Comparison with and without dynamical quasiparticle at different velocity 
  parameter for chiral susceptibility.}
  \label{fig:chim}
\end{center}
\end{figure}

To see the contribution of fermi velocity and vacuum polarization of bounded fermion 
into critical phenomena, we compare the peak position of chiral susceptibility at 
$v=0.1$ and $v=0.05$ (in addition to $v=0.03$ for full calculation) 
with and without dynamical fermion at mass parameter $m=0.0025$ in Figure \ref{fig:chim}. 
We notice two points:
\begin{enumerate}
\item First, our lattice calculation shows that critical point in full QED is moved to 
strong region in quenched QED $\beta v\simeq 0.093$ corresponding to $\alpha_c\sim 0.8$, 
which is about 40--50\% strongly shift of coupling constant. 
Our result is qualitatively 
consistent behavior with result in gap equation including frequent vacuum polarization 
effect of 2+1 dimensional fermion, quoted to be 46\% shift from 
$\alpha_c=0.5$ (quench) \cite{Wang:2010} to $\alpha_c=0.92$ \cite{Gamayun:2010}.
\item Secondly, decreasing the velocity from $v=0.1$ to $v=0.05$ we see
that the critical point $\alpha$ stays almost constant but shifts towards
stronger coupling for only about 5--7\% 
in both cases of full QED; peak position of $\chi_m$ is moved as 
$\beta v\simeq 0.062(v=0.1)\rightarrow 0.058(v=0.05)$, 
and quenched QED; peak position of $\chi_m$ is moved as 
$\beta v\simeq 0.093(v=0.1)\rightarrow 0.088(v=0.05)$.
Linearly extrapolating into $v\simeq 1/300$ using $\alpha_c(v) = \alpha_c(v=0) + cv$
assuming $c$ is constant, we find
that the critical point is $\alpha_{\rm c}(v=1/300)\sim 1.5$ 
in the case of full QED ($\alpha_{\rm c}(v=1/300)\sim 0.9$ in quenched QED),
which is comparable to TB model ($\alpha\sim 2.2$) \cite{Wallace:1947} 
rather than quenched one.
Assuming that the bare velocity $v$ and the renormalized velocity
$v_{\rm F}$ is almost the same, our results suggest 
the physical point of the Graphene system exists within the
symmetry broken phase.
To establish the position of the physical point in the phase structure, we
have to carry out the renormalization of the effective coupling constant as well as the
fermi velocity, which will be left a future work.
\end{enumerate}

\section{Discussion}\label{sec:discuss}
In this paper we have performed the lattice simulation using 4 dimensional 
non-compact QED with 2+1 dimensional fermion brane including velocity parameter 
into anisotropic gauge coupling between electric and magnetic fields. 
Our lattice simulation clearly shows that a consistent behavior associated with 
chiral symmetry breaking at critical coupling $\alpha_c=$1.45--1.59 at $v=0.1$, and 
furthermore from comparison with different velocity parameter 
we suggest possibility of quantum correction to the critical phenomena depending on 
running of velocity parameter in the infrared region. 
We assert that in our relativistic QED model related to 
``realistic'' Graphene system which includes the fermi-velocity  
less than 1\% speed of light and electromagnetic interaction among electron, 
the spontaneous chiral symmetry breaking occurs due to small velocity parameter, 
which plays important role to lead to the strong dynamics of this system rather than 
gauge coupling as suggested by perturbative argument \cite{Gonzalez:1994,Kotov:2010}.
Qualitatively the result obtained in our model is similar to 
model prediction of gap equation \cite{Gorbar:2002,Gamayun:2010}.
We argue that our model is non-trivial extension of QED model 
under gauge invariance to relativistic form 
in contrast to non-relativistic case \cite{Semenoff:1984,Drut:2009}.
Although we have not comprehended the explicit relation between 
our model and Graphene in low energy region yet, 
however surprisingly our model also shows 
the critical behavior of quasiparticle condensate 
as well as in the non-relativistic calculation 
\cite{Gonzalez:2010,Armour:2010,Drut:2009}.
This is strong evidence of that our model realizes the natural extension to 
relativistic theory of Graphene model beyond TB approximation.
We are now proceeding to take into account 
renormalization of velocity and continuum limit \cite{Note2} using more 
detailed parameter search under way.
This is a feasible study for one of the most challenging topics of theoretical 
prediction of semimetal-insulator transition point of suspended Graphene.

\begin{acknowledgments}
The calculations were performed by using the RIKEN Integrated Cluster of Clusters 
(RICC) facility. We also thank the support of computational resource from 
Texas Advanced Computing Center (TACC) for the disaster of earthquake in Japan
in 2011. This work is supported by the Grant-in-Aid of the Japanese Ministry of Education
(No. 20105002, 23105714).
\end{acknowledgments}


\begin{thebibliography}{99}
\bibitem{Note:1}
This statement is correct if the lattice cut-off in temporal direction 
does not depend on velocity parameter. Although it is necessary to perform careful scaling 
study, our rough estimate obtained in $v$ dependence of critical coupling
presented in Figure \ref{fig:chim} suggests that its scaling dependence 
is less than 20\% effect.
\end{thebibliography}
\end{document}